\pdfoutput=1
\documentclass[conference]{IEEEtran}
\IEEEoverridecommandlockouts
\usepackage{cite}
\usepackage{amsmath,amssymb,amsfonts}
\usepackage{graphicx}
\usepackage{booktabs}
\usepackage{url}
\usepackage{balance}
\usepackage{xcolor}
\usepackage{tikz}
\usetikzlibrary{arrows.meta,positioning,fit}
\graphicspath{{figures/}}
\makeatletter\g@addto@macro\normalsize{\setlength{\abovedisplayskip}{3pt plus 1pt}\setlength{\belowdisplayskip}{3pt plus 1pt}}\makeatother
\def\BibTeX{{\rm B\kern-.05em{\sc i\kern-.025em b}\kern-.08em T\kern-.1667em\lower.7ex\hbox{E}\kern-.125emX}}

\newcommand{\snrFull}{7.3}
\newcommand{\snrFilt}{1.9}

\newcommand{\emuSent}{347}
\newcommand{\emuBursts}{250}
\newcommand{\emuOverlap}{63}
\newcommand{\emuOneBeforePlain}{152}
\newcommand{\emuOneBeforeSplit}{171}
\newcommand{\emuOneAfterPlain}{90}
\newcommand{\emuOneAfterSplit}{157}
\newcommand{\emuTwoPlain}{194}
\newcommand{\emuTwoSplit}{279}
\newcommand{\emuTwoPlainPct}{56}
\newcommand{\emuTwoSplitPct}{80}

\newcommand{\fleetFourPlain}{27}
\newcommand{\fleetFourSplit}{61}
\newcommand{\fleetFourFree}{28}
\newcommand{\timePlain}{0.4}
\newcommand{\timeSplit}{2.7}

\begin{document}

\title{A Passive Software-Radio Ground Station for 433\,MHz UAV Telemetry by Bandpass Undersampling}

\author{\IEEEauthorblockN{Md Shahariar Islam Shanto}
\IEEEauthorblockA{\textit{Dept. of Engineering Technology}\\
\textit{Middle Tennessee State University}\\
Murfreesboro, TN, USA\\
ms2rt@mtmail.mtsu.edu}
\and
\IEEEauthorblockN{Tanzimul Alam Fahim}
\IEEEauthorblockA{\textit{Computational and Data Science}\\
\textit{Middle Tennessee State University}\\
Murfreesboro, TN, USA\\
mf9i@mtmail.mtsu.edu}
\and
\IEEEauthorblockN{Mohammad Arif Hossain}
\IEEEauthorblockA{\textit{Dept. of Engineering Technology}\\
\textit{Middle Tennessee State University}\\
Murfreesboro, TN, USA\\
mohammad.hossain@mtsu.edu}}

\IEEEaftertitletext{\vspace{-1.5\baselineskip}}

\maketitle

\begin{abstract}
Small unmanned aerial vehicles (UAVs) report their state over MAVLink through SiK-class telemetry radios, which pair two nodes in time division. In our measurements, adding a third radio to the network reduced the telemetry rate from 26 to 15 frames per second, and a fourth left only heartbeats. We study whether a receive-only software radio can observe such a link. Two USRP N210 units with BasicRX front ends rated to 250\,MHz receive the 433.92\,MHz link by bandpass undersampling at 100\,MS/s, and we recovered the link's PN9 whitening seed by exhaustive search against MAVLink's checksum. In an indoor corridor with a static UAV, the station recovered 98--99\% of the distinct frames that a native SiK radio delivered at 5 and 10\,m, and its reach with the unamplified front end ended between 15 and 20\,m, whereas the native link was unaffected out to 30\,m. The two MIMO-locked channels received the UAV with a coherence of 0.999. In simulation, the decoder achieves 90\% frame recovery at a burst-to-noise ratio of 7.3\,dB, or 1.9\,dB with a per-channel filter; in emulation with four UAVs sharing the band, it recovers 61\% of each UAV's frames.
\end{abstract}

\begin{IEEEkeywords}
UAV telemetry, MAVLink, SiK radio, software-defined radio, bandpass sampling, passive ground station
\end{IEEEkeywords}

\section{Introduction}
Research operations involving fleets of unmanned aerial vehicles (UAVs) are rarely managed from a single platform. Typically, the pilot's ground control station maintains the primary command-and-control link, a designated safety observer monitors safety-critical telemetry such as battery status and flight mode, a logging system records the full message stream for post hoc analysis, and experimental payloads or algorithms may require a low-latency live feed on an additional machine. Such shared access to telemetry becomes more important as UAVs take on roles as edge-computing platforms~\cite{hossain2022numerology}, multi-band aerial network nodes~\cite{hossain2023dual}, and components of satellite-ground integrated edge networks~\cite{liu2026multi}. In many ArduPilot and PX4 fleets, the telemetry link is a pair of SiK-class radios in the 433 or 915\,MHz band carrying MAVLink~\cite{koubaa2019mavlink,sik_firmware}. That link is designed for exactly two radios, and we measured how its capacity is shared when more are added: 26 frames per second with two radios, 15 with three, and heartbeats only with four (Section~\ref{sec:cost}). Because a fleet gives each UAV its own radio pair (Section~\ref{sec:fleet}), this measurement concerns a single pair and shows that an additional observer cannot join that pair as a radio without reducing its throughput. The usual alternative, forwarding MAVLink over the network from the machine that holds the radio, is inexpensive but ties every consumer to that machine; we did not compare the two approaches in cost or latency.

This paper asks whether a receive-only software-defined radio (SDR) can serve as such an observer while preserving the link's full capacity for the pilot's ground control station. Passive RF detection of drones is an active field~\cite{ezuma2020detection,nguyen2017drone}, and the openness of MAVLink is well documented~\cite{kwon2018mavlink,allouch2019mavsec}. The signal-processing components are standard, and the contribution of this work lies in integrating them into a working station built on generic hardware, recovering the radio parameters, and measuring the station's performance. The receivers are two Ettus USRP N210 units with BasicRX daughterboards, which have no mixer, amplifier, or filter and are rated for 1--250\,MHz, while the telemetry is transmitted at 433.92\,MHz. We receive it by bandpass undersampling~\cite{vaughan1991bandpass}: the N210 samples at 100\,MS/s, and because the converter's input bandwidth extends well beyond the first Nyquist zone, the signal folds to 33.92\,MHz. These radios were chosen because they were available in the laboratory and because the MIMO cable provides two coherent channels; this is a feasibility study, not a comparison of front ends.

The contributions are: (i) a measured demonstration that a direct-sampling front end rated to 250\,MHz receives a 433.92\,MHz frequency-hopping GFSK link, with no reception at 2.4\,GHz and 915\,MHz untested; (ii) recovery of the SiK link's whitening seed by exhaustive search with MAVLink's checksum as the oracle, reported here so that others need not repeat the search; (iii) a decoder that converts the captures into the telemetry a dashboard displays, measured against the native radio on an indoor distance ladder and in repeated runs; (iv) measured two-channel coherence as a basis for a bearing mode, which is not itself demonstrated here; (v) a system model with the equations that describe each stage; and (vi) simulation and emulation, explicitly labeled as such, that characterize the decoder's sensitivity and its behavior with several UAVs on a single band.

\section{Background and Related Work}
The SiK firmware runs on Si1000-class transceivers (HM-TRP, RFD900, and similar) and implements a two-node frequency-hopping link with a fixed time-division window, optional Golay coding, and a transparent serial interface~\cite{sik_firmware,si4432_datasheet}. MAVLink frames carried over this link end with a 16-bit CRC extended by a per-message constant, so any receiver can verify unambiguously that a candidate frame is genuine~\cite{mavlink_spec}. Koub\^aa et al.~\cite{koubaa2019mavlink} survey the protocol, while Kwon et al.~\cite{kwon2018mavlink} and Allouch et al.~\cite{allouch2019mavsec} show that anyone able to read it can also inject into it, and they propose signing and encryption as countermeasures. The Si443x preamble, sync word 0x2DD4, and packet structure are documented~\cite{si4432_datasheet}, and the SiK settings can be read from the radios, so only the whitening state at the first payload byte remained to be determined. We use MAVLink's openness as the oracle for that search.

Bandpass sampling maps a signal sampled below its Nyquist rate into the first Nyquist zone, provided that the band does not straddle a multiple of $f_s/2$~\cite{vaughan1991bandpass}; the practical constraints are the converter's input bandwidth and the folding of every other zone onto the same image. Passive RF sensing of drones detects and classifies their control links by energy and fingerprint~\cite{ezuma2020detection,nguyen2017drone}, and sensing is increasingly integrated with communication in 6G network design~\cite{hossain2023ai}, whereas the present work decodes the link, which turns the observer into a telemetry consumer. Software receivers for Si443x-class packet radios exist in the hobbyist community, but we found no published, measured account of a SiK/MAVLink observer.

\section{Ground Station Architecture}
\label{sec:arch}
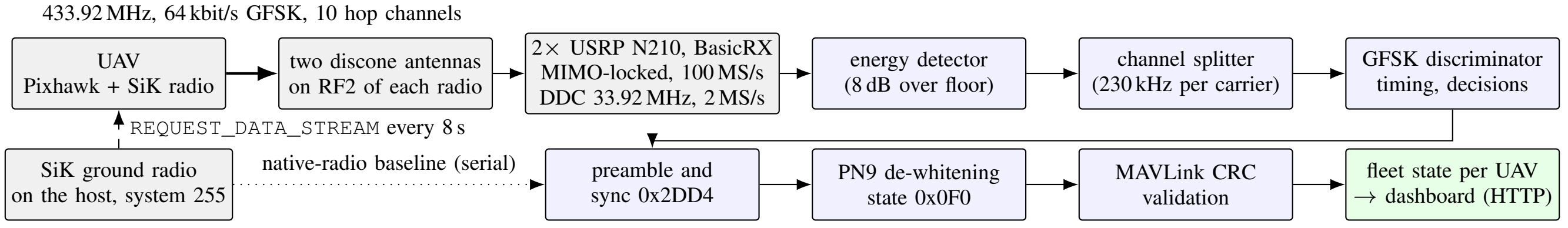
\begin{figure*}[t]
\centering
\begin{tikzpicture}[font=\scriptsize, >=Latex,
  box/.style={draw, rounded corners=1pt, minimum height=8mm, minimum width=24mm, align=center, inner sep=2pt},
  stage/.style={box, fill=blue!6},
  hw/.style={box, fill=gray!12},
  res/.style={box, fill=green!10}]
\node[hw] (ac) at (0,0) {UAV\\Pixhawk + SiK radio};
\node[hw] (ant) at (3.0,0) {two discone antennas\\on RF2 of each radio};
\node[hw] (n210) at (6.0,0) {2$\times$ USRP N210, BasicRX\\MIMO-locked, 100\,MS/s\\DDC 33.92\,MHz, 2\,MS/s};
\node[stage] (det) at (9.0,0) {energy detector\\(8\,dB over floor)};
\node[stage] (chan) at (12.0,0) {channel splitter\\(230\,kHz per carrier)};
\node[stage] (dem) at (15.0,0) {GFSK discriminator\\timing, decisions};
\node[hw] (sik) at (0,-1.25) {SiK ground radio\\on the host, system 255};
\node[stage] (sync) at (6.0,-1.25) {preamble and\\sync 0x2DD4};
\node[stage] (wh) at (9.0,-1.25) {PN9 de-whitening\\state 0x0F0};
\node[stage] (crc) at (12.0,-1.25) {MAVLink CRC\\validation};
\node[res] (state) at (15.0,-1.25) {fleet state per UAV\\$\rightarrow$ dashboard (HTTP)};
\draw[->, thick] (ac) -- node[above=4.2mm, font=\scriptsize]{433.92\,MHz, 64\,kbit/s GFSK, 10 hop channels} (ant);
\draw[->] (ant) -- (n210); \draw[->] (n210) -- (det); \draw[->] (det) -- (chan); \draw[->] (chan) -- (dem);
\draw[->] (dem.south) -- ++(0,-0.35) -| (sync.north);
\draw[->] (sync) -- (wh); \draw[->] (wh) -- (crc); \draw[->] (crc) -- (state);
\draw[->, dashed] (sik) -- node[right, align=left]{\texttt{REQUEST\_DATA\_STREAM} every 8\,s} (ac);
\draw[->, dotted] (sik.east) -- node[above]{native-radio baseline (serial)} (sync.west);
\end{tikzpicture}
\caption{The passive ground station. Solid: the receive path, which transmits nothing. Dashed: the one transmission in the system, a stream request from a separate SiK radio. Dotted: that radio's serial output, the native-radio baseline.}
\label{fig:arch}
\end{figure*}

Figure~\ref{fig:arch} shows the station. Two discone antennas feed the BasicRX inputs of the two N210 units, and the second N210 receives its 100\,MHz clock and time from the first through the MIMO expansion cable, so both channels are sampled on the same clock. Each unit's digital down-converter (DDC) is tuned to the 33.92\,MHz image and delivers a 2\,MS/s complex stream to the host, which covers the entire 433.05--434.79\,MHz hopping band. The host runs the decoder of Section~\ref{sec:model} in Python with NumPy and SciPy, maintains a state record for each UAV (mode, battery, position, attitude, status text, and link statistics), and serves that record as JSON to a browser dashboard on the laboratory network. The only transmission in the system is a \texttt{REQUEST\_DATA\_STREAM} message sent every 8\,s by a separate SiK radio on the host as system 255, because ArduPilot streams attitude, battery, and position only when a ground station requests them. The station therefore requires a requesting ground station on the link, since without the request the UAV sends only heartbeats, and the same SiK radio also serves as the native-radio baseline.

\section{System Model}
\label{sec:model}

\subsection{Undersampled Reception}
The station does not translate the 433.92\,MHz signal down in frequency before digitization; instead, the converter samples it directly at a rate far below its Nyquist rate and relies on aliasing to bring it into the first Nyquist zone. Under this bandpass sampling, a carrier $f_c$ sampled at $f_s$ appears at
\begin{equation}
f_{\mathrm{img}} = \bigl|f_c - m f_s\bigr|,\qquad m=\operatorname{round}(f_c/f_s),
\label{eq:alias}
\end{equation}
in Nyquist zone $k = \lfloor 2f_c/f_s\rfloor + 1$, and it is spectrally inverted when $k$ is even. With $f_c = 433.92$\,MHz and $f_s = 100$\,MS/s, we have $m=4$, $f_{\mathrm{img}} = 33.92$\,MHz, and $k = 9$, so the image is upright and the entire hop set lies within a 2\,MHz band around it. The signal reaches the converter through the BasicRX transformer and the converter's own sample-and-hold, whose full-power bandwidth $f_{3\mathrm{dB}}$ (575\,MHz for the N210's LTC2284) attenuates it by approximately
\begin{equation}
G(f_c) = -10\log_{10}\bigl[1+(f_c/f_{3\mathrm{dB}})^2\bigr] \approx -2.0\ \mathrm{dB}.
\label{eq:rolloff}
\end{equation}
This figure covers the sample-and-hold only and excludes the loss of the BasicRX transformer at 434\,MHz, which we did not measure. Without a preselector, signals and noise from every other Nyquist zone fold onto the same image, and we did not quantify this contribution with a bandpass filter.

\subsection{Burst Detection and Demodulation}
The transmitted burst is 2-GFSK,
\begin{equation}
s(t) = \exp\Bigl\{ j2\pi \Delta f \!\int_{-\infty}^{t}\! \sum_k a_k\, g(\tau - kT_b)\,d\tau \Bigr\},\quad a_k \in\{\pm1\},
\label{eq:gfsk}
\end{equation}
with $\Delta f = 66.6$\,kHz, $T_b = 1/64$\,ms, and $g$ a unit-height rectangular pulse of width $T_b$ smoothed by a unit-area Gaussian, which gives a modulation index $h = 2\Delta f T_b \approx 2.1$. Bursts are detected by energy: the envelope in any 100\,$\mu$s block must exceed the capture's 25th percentile by 8\,dB. The 25th percentile lies on the noise floor whenever the band is idle for more than a quarter of the time, which holds for these captures. Within a burst, the carrier offset $\hat f_0$ is estimated as the power centroid of the burst spectrum, and after this offset is removed, the frequency discriminator
\begin{equation}
\hat f[n] = \frac{f_s'}{2\pi}\arg\bigl(x[n]\,x^{*}[n-1]\bigr),\qquad f_s' = 2\ \mathrm{MS/s},
\label{eq:disc}
\end{equation}
is low-pass filtered to $1.2/T_b$. The symbol timing $\hat\tau$ maximizes $\sum_k |\hat f(kT_b+\hat\tau)|$ over one symbol of 31.25 samples, and the bit decision is $\hat a_k = \operatorname{sgn}\hat f(kT_b+\hat\tau)$. The channel-splitting variant, used alongside the plain decoder in Sections~\ref{sec:meas}--\ref{sec:fleet}, inserts one step before (\ref{eq:disc}): it locates the carriers occupied in the burst's smoothed spectrum, mixes each of them to baseband, and filters each to a 230\,kHz channel before demodulating them separately.

\subsection{Whitening Recovery as a Search}
After the 66-bit preamble and the sync word, the transceiver whitens the bytes with a 9-bit linear-feedback shift register (LFSR). With state $r$ and taps at bits 0 and 5, the update is
\begin{equation}
r \leftarrow (r \gg 1)\ \lor\ \bigl[(r \oplus (r \gg 5)) \wedge 1\bigr] \ll 8,
\label{eq:pn9}
\end{equation}
where each step yields one bit of the whitening byte $w_i$, assembled most significant bit first, and the clear byte is $d_i = c_i \oplus w_i$. The polynomial is standard, whereas the tap positions, bit order, and initial state $r_0$ are implementation choices. We score every hypothesis $(r_0, \text{taps}, \text{order})$, $512\times2\times2 = 2048$ in all, by
\begin{equation}
S(r_0,\cdot) = \sum_{\text{bursts}} \#\{\text{CRC-valid MAVLink frames in } d\},
\label{eq:score}
\end{equation}
and keep the maximizer; the full search takes about one second on an 8\,s capture.

\subsection{The Validation Oracle}
A MAVLink frame validates when its CRC-16/MCRF4XX over the header and payload, extended by the message's constant, matches the trailing two bytes. A candidate at a random offset matches with probability $2^{-16}$, so a wrong whitening state yields almost no valid frames: in our capture, no wrong state produced more than 6, against 159 for the correct state (Section~\ref{sec:phy}).

\section{Testbed}
\label{sec:testbed}
\begin{figure*}[t]
\centering
\includegraphics[width=0.65\textwidth]{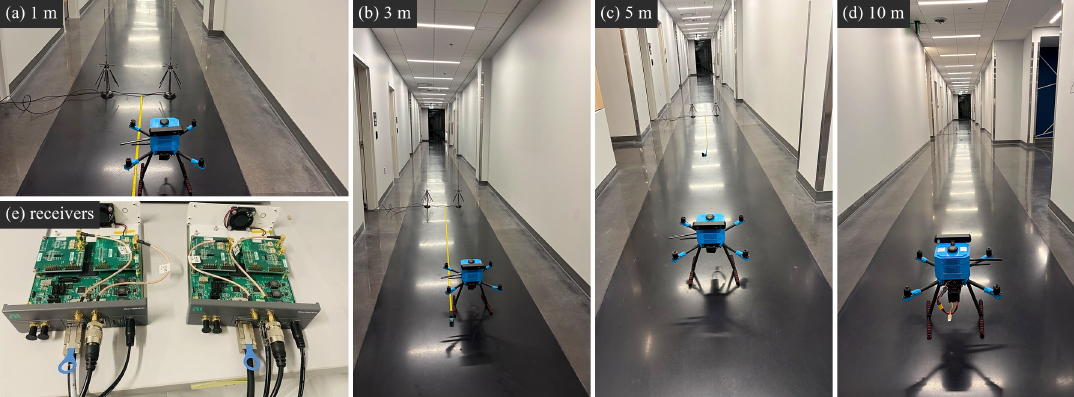}
\caption{The testbed. (a)--(d) The corridor ladder: two discone antennas on the floor 30\,cm apart, with the UAV on the tape at 1, 3, 5, and 10\,m. (e) The two USRP N210 units with lids removed, showing the BasicTX and BasicRX boards, the MIMO cable between them (blue tab), and the antennas on RF2.}
\label{fig:testbed}
\end{figure*}

Figure~\ref{fig:testbed} shows the equipment. The receivers are two USRP N210 (rev 4) units, each with one BasicTX and one BasicRX daughterboard, running firmware 12.4 and FPGA image 11.1 under UHD 4.6 on a Linux laptop with a dedicated gigabit Ethernet port. The UAVs are three Holybro X500-class quadcopters, each with a Pixhawk running ArduCopter, a Jetson companion computer, and an RFD SiK 2.0 radio on the flight controller's TELEM port. The radio settings, as read over the AT interface, are: air speed 64\,kbit/s, 10 channels within the band limits of 433\,050--434\,790\,kHz (carriers measured at 433.16--434.51\,MHz), 20\,dBm output power, network ID 25, MAVLink framing on, a 131\,ms window, and error correction, Manchester coding, and listen-before-talk all off. The decoder relies on the air speed, the absence of coding, and the band plan.

\section{Recovering the Air Interface}
\label{sec:phy}
Tuning to the image of (\ref{eq:alias}) reveals the link directly: bursts of 2.5 to 28\,ms on ten channels spaced 150\,kHz apart between 433.16 and 434.51\,MHz (Fig.~\ref{fig:frontend}, left). With the UAV about 1\,m from the antenna, the bursts have a median level of $-55.9$\,dBFS against a receiver floor of $-80.8$\,dBFS (Section~\ref{sec:antenna}), while the ground radio's own packets, transmitted next to the antenna, arrive at $-27$\,dBFS. With a 2.4\,GHz antenna on the same input and the UAV's Wi-Fi transmitting beside the bench, the level stayed within 1.2\,dB of the floor at every image we tried; the front end therefore received nothing at 2.4\,GHz, and frequencies between 434\,MHz and 2.4\,GHz were not tested.

\begin{figure}[t]
\centering
\includegraphics[width=0.65\columnwidth]{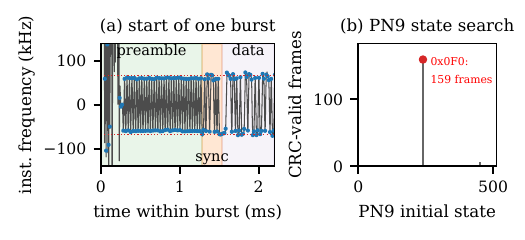}
\caption{(a) Instantaneous frequency (\ref{eq:disc}) at the start of one received burst with the decoder's symbol decisions: preamble, sync word, and the start of the whitened payload at $\pm66.6$\,kHz. (b) The search of (\ref{eq:score}) over the 512 initial states. Measured.}
\label{fig:anatomy}
\end{figure}

Table~\ref{tab:phy} lists the air interface as recovered from the captures, and Fig.~\ref{fig:anatomy}(a) shows the start of one received burst with the decoder's symbol decisions. The whitening search of (\ref{eq:score}), shown in Fig.~\ref{fig:anatomy}(b) for the winning taps and bit order, is unambiguous: state 0x0F0 yields 159 CRC-valid frames from all senders in the 8\,s capture (counted with the plain decoder), every other state yields no more than 6, and both the unwhitened bytes and the LSB-first variant yield none. PN9 whitening normally starts from 0x1FF, and we do not know whether 0x0F0 is that state advanced to the first byte after the sync word or a different seed. The decoder found the sync word in 247 of the 258 bursts in the capture. Of the 11 misses, 4 were shorter than 2\,ms and 7 were among the 254 longer bursts, which gives a detection rate of 97\% for bursts longer than 2\,ms. Of all the bursts, 51 carried MAVLink from the UAV, and the remainder were the radios' own keep-alive packets, which carry no payload and are not errors.

\begin{table}[t]
\centering
\caption{Air interface recovered from captures}
\label{tab:phy}
\scriptsize
\setlength{\tabcolsep}{3.5pt}
\begin{tabular}{lll}
\toprule
Property & Value & How it was found \\
\midrule
Modulation & 2-GFSK & spectrum, frequency trace \\
Symbol rate & 64.0\,kbit/s & transition spacing in the trace \\
Deviation & $\pm$66.6\,kHz ($h \approx 2.1$) & frequency-trace levels \\
Preamble & 66 bits, 1010\dots & frequency trace \\
Sync word & 0x2DD4 & bit pattern after preamble \\
Whitening & PN9, $x^9+x^5+1$, MSB first & search, (\ref{eq:score}) \\
Initial state & 0x0F0, from the first byte & search, (\ref{eq:score}) \\
Link layer & header, payload, 16-bit CRC & MAVLink CRC oracle \\
Hop set & 10 channels, 150\,kHz apart & burst carrier estimates \\
\bottomrule
\end{tabular}
\end{table}

\section{Measurements}
\label{sec:meas}

\subsection{What the Station Shows}
\begin{figure}[t]
\centering
\includegraphics[width=0.60\columnwidth]{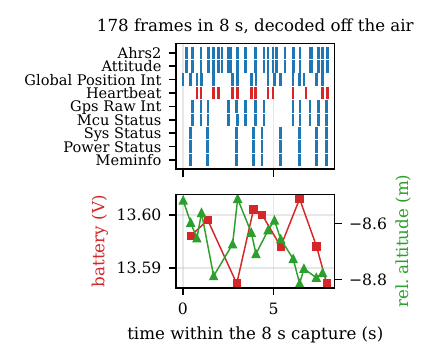}
\caption{The 8\,s capture decoded. Top: arrival time of every frame by message type (heartbeats in red). Bottom: battery voltage and relative altitude, as shown on the dashboard. Measured with the UAV on the bench.}
\label{fig:telemetry}
\end{figure}

Figure~\ref{fig:telemetry} shows the 8\,s capture decoded into the quantities that the dashboard displays. With the channel-splitting decoder, the station recovered 171 frames from the UAV and 7 from its companion computer, carrying attitude, AHRS, position, GPS, MCU status, battery, heartbeat, and status text.

\subsection{The Cost of Joining the Link}
\label{sec:cost}
With two radios on the link (the UAV and the ground radio), the laptop's SiK radio received 26 frames per second (7.0\,kbit/s). With a third SiK radio on the same network ID, powered and within range, it received 15 frames per second (4.1\,kbit/s), a reduction of 42\%. With a fourth radio, the rate fell to about one frame per second (heartbeats only), which we attribute to the pair losing time-division synchronization. With the passive station present instead of the additional radios, the rate stayed at 26 frames per second, so the station imposes no measurable cost on the link.

\begin{figure*}[t]
\centering
\includegraphics[width=0.75\textwidth]{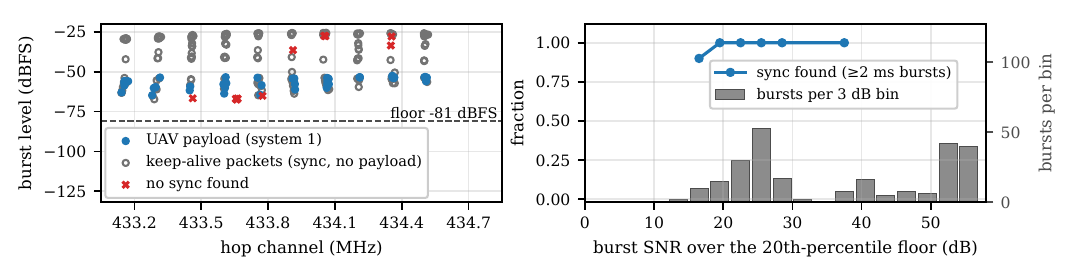}
\caption{Front end on the bench capture. Left: the level of every burst against its hop channel. Right: fraction of bursts whose sync word was found against SNR over the 20th-percentile floor, with bursts per bin as bars. Measured. The simulation, which uses the mean noise power instead, is shown in Fig.~\ref{fig:sens}.}
\label{fig:frontend}
\end{figure*}

\subsection{What the Station Recovers: Distance Ladder and Repeats}
\label{sec:diversity}
\begin{figure}[t]
\centering
\includegraphics[width=0.65\columnwidth]{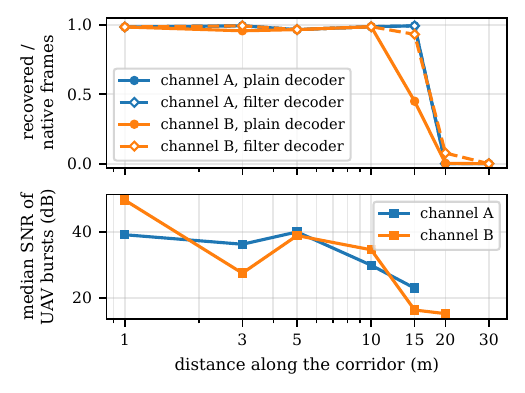}
\caption{Corridor ladder. Top: frames recovered by the station over frames delivered by the native radio in the same 15\,s, for both channels and both decoders. Bottom: median SNR of the decoded UAV bursts over the 20th-percentile floor. Measured.}
\label{fig:ladder}
\end{figure}

The station was compared with the native radio in a 30\,m corridor (Fig.~\ref{fig:testbed}(a)--(d)), with the antennas on the floor 30\,cm apart, the UAV on a tape measure at 1, 3, 5, 10, 15, 20, and 30\,m, 15\,s per point, the laptop's SiK radio next to the antennas as the native receiver, and one UAV powered. Fig.~\ref{fig:ladder} shows the ratio of frames recovered by the station to frames delivered by the native radio in the same window for both channels and both decoders, with the SNR of the decoded UAV bursts below. Out to 10\,m, the station recovered 96--99\% on both channels. At 15\,m, channel A still recovered 99\%, whereas channel B, which was 7\,dB weaker at that point, dropped to 45\% with the plain decoder and recovered 93\% (439 of 471) with the channel filter. At 20\,m, the energy detector picked up only a third of the bursts, the plain decoder recovered a single frame, and the channel filter recovered 31 of 404. At 30\,m, the station recovered nothing, while the native radio delivered 27--32 frames per second at every point. The 15--20\,m edge is set by this corridor, the floor-level antennas, and the unamplified front end, and it would differ at another site or with an amplifier. Three further 20\,s runs at 5\,m gave 625, 626, and 625 frames against 576, 616, and 632 at the native radio, a ratio of $103 \pm 5\%$ (mean and sample standard deviation over the three runs, with identical counts on both channels), which shows consistent short-range performance under these conditions but does not establish general equivalence. A ratio above 100\% mixes recall with frames that the native radio missed, so we report both quantities separately. MAVLink numbers each sender's frames modulo 256, so with received sequence numbers $s_j$, the fraction of frames that a receiver missed is
\begin{equation}
L = \frac{\sum_j \bigl[\bigl((s_{j+1}-s_j) \bmod 256\bigr) - 1\bigr]}{\;\text{received} + \text{missed}\;},
\label{eq:seqloss}
\end{equation}
computed per sender (system and component), where a gap of exactly 256 frames is invisible to this measure. By (\ref{eq:seqloss}), the native radio missed 50 of 609 UAV frames (8\%) in the first repeat and none at the 5\,m ladder point. Keyed on sender, sequence number, and message ID, the station missed 8 of the 466 distinct frames that the native radio delivered at 5\,m (98.3\% recall) and 6 of the 472 at 10\,m (98.7\% recall). In two of the repeats, the station recovered 21 and 53 frames, respectively, that the native radio had missed. The decoder silently drops any message whose CRC constant is absent from its table; the table used here includes AHRS2 and MCU\_STATUS, which an earlier version of the decoder lacked.

\subsection{Antenna and Front End at 433\,MHz}
\label{sec:antenna}
Figure~\ref{fig:frontend} characterizes the receive chain burst by burst on the bench capture. The UAV payload bursts have a median level of $-55.9$\,dBFS across all ten channels, with no channel deviating noticeably, as expected over a fractional bandwidth of 0.4\%. The larger population between $-30$ and $-45$\,dBFS is the keep-alive traffic, consistent with the ground radio's position next to the antenna. The receiver floor, estimated as the 20th percentile of sample power over the capture with the same radios, sample rate, and processing (the BasicRX has no gain setting), is $-80.8$\,dBFS on the bench and $-90.3$\,dBFS in the corridor. We attribute the difference to pickup from the laptop, radios, and power supplies next to the bench antennas, which in the corridor were 2\,m away and produced no pickup. The UAV bursts therefore arrive at an SNR of 17--35\,dB on the bench and 30--50\,dB at 1--10\,m in the corridor.
The sync word was present in 97\% of the bursts longer than 2\,ms (247 of 254, $\pm2\%$ binomial) and in 100\% of those between 20 and 40\,dB. Finding the sync word is a weaker test than decoding a full frame, and all 51 UAV payload bursts passed both. The 11 misses are the weakest UAV bursts, near 16\,dB, as expected given the receiver sensitivity, and are accompanied by a few of the strongest keep-alive bursts, possibly due to clipping or overlapping bursts, which we did not investigate.

\subsection{Coherent Two-Channel Reception}
\label{sec:coherent}
On the UAV bursts received through the two matched discone antennas, the complex correlation between the two MIMO-locked channels was 0.999, the burst-to-burst standard deviation of their phase difference $\sigma_\phi$ was $1.2^{\circ}$ in one session and $1.7^{\circ}$ in another, and their levels were within 1.2\,dB of each other. For antennas spaced $d$ apart, the phase difference is $\phi = 2\pi d\sin\theta/\lambda$~\cite{poisel2012ew}, so the bearing precision is
\begin{equation}
\sigma_\theta = \frac{\lambda\,\sigma_\phi}{2\pi d\cos\theta},
\label{eq:bearing}
\end{equation}
which at $\lambda = 0.691$\,m and $d = 0.30$\,m (inside the half-wavelength spacing that keeps the estimate unambiguous) gives $0.44^{\circ}$ at boresight for $\sigma_\phi = 1.2^{\circ}$. The fixed phase offset between the radios is arbitrary at each power-up ($+84^{\circ}$ in one session and $-120^{\circ}$ in another) and is calibrated against a known bearing. On the UAV's own bursts along the corridor ladder, with the UAV on the axis throughout, $\sigma_\phi$ within a capture was $3$--$6^{\circ}$ at 1, 5, and 10\,m with a coherence of 0.999, which by (\ref{eq:bearing}) corresponds to $1.1$--$2.2^{\circ}$ of bearing per burst, and the three repeats at 5\,m agreed in mean within $1.5^{\circ}$ ($0.55^{\circ}$ of bearing). Between positions on the same axis, however, the mean phase shifted by up to $90^{\circ}$ (about $35^{\circ}$ of apparent bearing), and at 3 and 15\,m, where one channel was in a fade, $\sigma_\phi$ widened to $25$--$45^{\circ}$. This shift is a multipath bias rather than noise, so a single calibration does not absorb it and averaging does not remove it. Coherent reception is therefore demonstrated, but direction finding is not; bearing estimates against known angles and against the GPS position that the UAV reports over the same link are the next measurements.

\section{Simulation: Sensitivity of the Decoder}
\label{sec:sim}
\begin{figure}[!t]
\centering
\includegraphics[width=0.65\columnwidth]{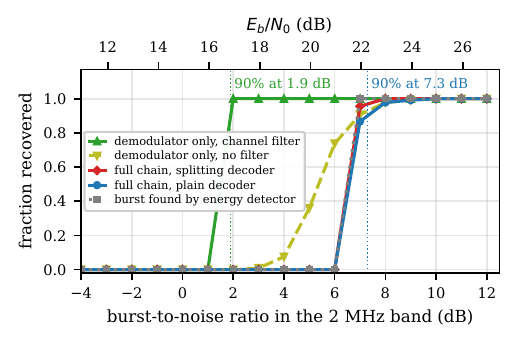}
\caption{Simulated frame recovery of the real decoder against burst-to-noise ratio in the 2\,MHz band (bottom) and $E_b/N_0$ (top), with and without the channel filter and the energy detector. Simulation.}
\label{fig:sens}
\end{figure}

Synthetic bursts following Table~\ref{tab:phy}, containing real MAVLink frames with valid checksums, were placed at random times and carrier offsets (within $\pm30$\,kHz) in white noise and passed through the same decoder used on the captures, with 150 bursts per point (Fig.~\ref{fig:sens}). The decoder as used in the measurements reaches 90\% recovery at a burst-to-noise ratio of \snrFull\,dB in the 2\,MHz band ($E_b/N_0 \approx 22$\,dB). Supplying the demodulator with the true burst position changes the result very little, so the energy detector does not set the limit for this chain. Placing the 230\,kHz channel filter before (\ref{eq:disc}) moves the 90\% point of the demodulator to \snrFilt\,dB; with the filter in place, however, the wideband energy detector becomes the limiting stage at about 7\,dB, and a preamble correlator would be the natural next step. The ladder shows why free-space margins must not be extrapolated indoors: the SNR of the decoded bursts fell from 30--35\,dB at 10\,m to 16--23\,dB at 15\,m, and the detector lost most bursts by 20\,m, a drop several times steeper than free space, which we attribute to the corridor's doorways and the floor-level antennas. The ladder agrees with the simulation in trend only: at the edge, the plain decoder fails first, the channel filter recovers a large part of what it loses (channel B at 15\,m rose from 45\% to 93\%), and the energy detector then sets the limit. The absolute thresholds do not transfer, since channel B failed with the plain decoder at a measured 16\,dB compared with the simulated \snrFull\,dB. About 6.5\,dB of that gap comes from the noise reference (the 20th percentile of sample power in the measurements against the mean noise power in the simulation); the remainder is unexplained, and the ladder does not confirm the simulated 5\,dB gain of the filter.

\section{Emulation: Several UAVs on One Band}
\label{sec:fleet}
\begin{figure}[t]
\centering
\includegraphics[width=0.65\columnwidth]{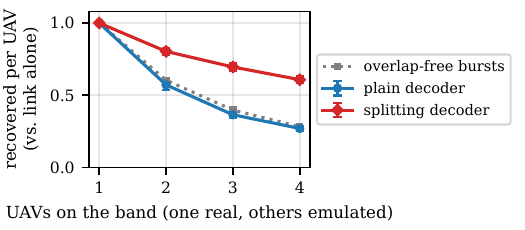}
\caption{Per-UAV frame recovery, relative to each link alone, against the number of UAVs sharing the band: one real UAV plus one to three emulated links added to the real capture, with the mean and standard deviation over ten random schedules. Emulation.}
\label{fig:fleet}
\end{figure}

A fleet gives each UAV its own radio pair and network ID, all hopping over the same band on independent schedules, so a ground radio sees only its own pair, whereas the station sees the whole band. On the real capture alone, the channel-splitting decoder recovers \emuOneBeforeSplit{} UAV frames against \emuOneBeforePlain{} for the plain decoder, a 13\% gain from the channel filter, and both decode faster than real time (\timePlain\,s and \timeSplit\,s for the 8\,s capture in single-threaded Python). We then emulated one, two, and three additional UAVs by generating their links at the real link's burst rate and median level ($\pm3$\,dB), hopping at random over the ten channels, adding them at the IQ level to the real 8\,s capture, and running both decoders. With one added UAV, \emuOverlap{} of its \emuBursts{} bursts overlapped a real burst in time. Because the plain decoder treats each energy burst as a single transmission, overlapping bursts on different channels share one carrier estimate, and it therefore kept only 59\% of the real UAV's frames (\emuOneAfterPlain{} of \emuOneBeforePlain) and recovered \emuTwoPlainPct\% of the added UAV's frames (\emuTwoPlain{} of \emuSent). The channel-splitting decoder kept 92\% of the real UAV's frames (\emuOneAfterSplit{} of \emuOneBeforeSplit) and recovered \emuTwoSplitPct\% of the added UAV's frames (\emuTwoSplit{} of \emuSent), losing only the bursts that overlap in both time and channel. Fig.~\ref{fig:fleet} extends this to four UAVs on the band, with ten random schedules each. With four UAVs, only \fleetFourFree\% of bursts are free of any time overlap, and the plain decoder tracks that figure at \fleetFourPlain\% per UAV, whereas the splitting decoder recovers \fleetFourSplit\%, since a burst that overlaps another on a different channel remains readable once the channels are separated. This \fleetFourSplit\% should be read against the near-complete reception of a dedicated pair, and it rests on optimistic assumptions: equal received power ($\pm3$\,dB) rather than the 20\,dB near-far spread of a real fleet, and adjacent-channel leakage that the emulation does not stress.

\section{Conclusion}
In an indoor feasibility study, a general-purpose SDR with a front end rated to 250\,MHz received a 433.92\,MHz UAV telemetry link via bandpass undersampling without participating in the link. The station decoded 98--99\% of the distinct frames that the native radio delivered from a static UAV at 5 and 10\,m, and its range with the unamplified front end ended between 15 and 20\,m. The two MIMO-locked channels received the UAV coherently, and simulation and emulation characterized the decoder's sensitivity and behavior with several UAVs on a single band. These results apply to a static UAV indoors and do not predict the range or performance achievable outdoors or in flight, and neither bearing estimation nor fleet operation has been demonstrated. Future work will repeat the measurements outdoors and in flight under a range of conditions and will evaluate bearing estimation and multi-UAV reception against ground truth in real deployments.

\bibliographystyle{IEEEtran}
\bibliography{references}

\end{document}